\documentclass[prd,english,preprintnumbers,amsmath,amssymb,nofootinbib,superscriptaddress,twocolumn]{revtex4-1}
\pdfoutput=1

\usepackage{mathtools}
\usepackage{amsfonts}
\usepackage{mathrsfs}
\usepackage{bbm}
\usepackage{babel}

\usepackage{graphicx}
\usepackage[dvipsnames]{xcolor}
\usepackage{array}

\usepackage{xspace}
\usepackage{siunitx}
\usepackage{hyperref}

\usepackage{xifthen}
\usepackage{dsfont}
\usepackage{appendix}
\usepackage{enumitem}
\usepackage{booktabs}
\usepackage{units}

\usepackage[latin1]{inputenc}
\usepackage[dvipsnames]{xcolor}
\usepackage{array}

\setkeys{Gin}{width=0.48\textwidth}

\graphicspath{
	{./}
}

\def\0#1#2{\frac{#1}{#2}}

\def\s0#1#2{\mbox{\small{$ \frac{#1}{#2} $}}}

\def\CD{{\mathcal D}}
\def\CZ{{\mathcal Z}}

\newcommand{\be}{\begin{eqnarray}}
\newcommand{\ee}{\end{eqnarray}}

\newcommand{\nn}{\nonumber }

\newcommand{\beq}{\begin{equation}}
\newcommand{\eeq}{\end{equation}}
\newcommand{\bea}{\begin{eqnarray}}
\newcommand{\eea}{\end{eqnarray}}

\newcommand{\up}{\uparrow}
\newcommand{\down}{\downarrow}

\def\0#1#2{\frac{#1}{#2}}

\hypersetup{
	colorlinks,
	linkcolor={red!75!black},
	citecolor={blue!75!black},
	urlcolor={blue!75!black},
	pdftitle={Pairing patterns in polarized unitary Fermi gases above the superfluid transition},
	pdfauthor={Attanasio, Rammelmueller, Drut, Braun},
	pdfkeywords={Complex Langevin, Correlation functions, Unitary Fermi gas},
	bookmarksopen=true,
	bookmarksopenlevel=2,
	bookmarksnumbered=true
}

\makeatother
\begin{document}

\title{Pairing patterns in polarized unitary Fermi gases above the superfluid transition}

\author{Felipe Attanasio}
\affiliation{Institut f\"ur Theoretische Physik, Universit\"at Heidelberg, Philosophenweg 16, 69120 Heidelberg, Germany}

\author{Lukas Rammelm\"uller}
\affiliation{Arnold Sommerfeld Center for Theoretical Physics (ASC), University of Munich, Theresienstr. 37, 80333 M\"unchen, Germany}
\affiliation{Munich Center for Quantum Science and Technology (MCQST), Schellingstr. 4, 80799 M\"unchen, Germany}

\author{Joaqu\'in E. Drut}
\affiliation{Department of Physics and Astronomy, University of North Carolina, Chapel Hill, North Carolina 27599, USA}

\author{Jens Braun}
\affiliation{Institut f\"ur Kernphysik (Theoriezentrum), Technische Universit\"at Darmstadt, D-64289 Darmstadt, Germany}
\affiliation{ExtreMe Matter Institute EMMI, GSI, Planckstra{\ss}e 1, D-64291 Darmstadt, Germany}

\begin{abstract}
We non-perturbatively study pairing in the high-temperature regime of polarized unitary two-component Fermi gases by extracting the 
pair-momentum distribution and shot-noise correlations. Whereas the pair-momentum distribution allows us to analyze the propagation of pairs composed 
of one spin-up and one spin-down fermion, 
shot-noise correlations provide us with a tomographic insight into pairing correlations
around the Fermi surfaces associated with the two species. Assuming that the dominant pairing patterns right above the superfluid transition also govern the formation of 
condensates in the low-temperature regime, our
analysis suggests that the superfluid ground state is homogeneous and of the Bardeen-Cooper-Schrieffer-type over a wide range of polarizations.
\end{abstract}

\maketitle
%
{\it Introduction.--}
In spite of the substantial progress in the field of ultracold atoms, both on the theoretical, computational, and experimental fronts,
the question of the existence of an inhomogeneous superfluid phase in strongly coupled Fermi gases
at low temperatures remains an open and challenging area of research. One of the most sought after cases is the
unitary limit of nonrelativistic spin-1/2 fermions, where the attractive zero-range interaction is tuned to resonance and the system is scale invariant.
At low enough temperatures, the unpolarized system displays a superfluid phase which, as the polarization is increased,
eventually disappears at some temperature-dependent critical polarization~\cite{Zwierlein27012006,*ZwierleinNature2006,*PhysRevLett.97.030401,*Shin2008,*Schunck867,Partridge27012006,*PhysRevLett.97.190407,PhysRevA.88.063614,PhysRevLett.96.060401,Gubbels:2006zz,2007NatPh...3..124P,Gubbels:2007xc,Boettcher2014,Roscher:2015xha,Frank_2018}, see Refs.~\cite{Chevy2010,Radzihovsky2010,Zwerger2012,GUBBELS2013255} for reviews.
Exactly how this happens, i.e., what exotic superfluid phases are traversed as the polarization is increased and to what extent they are stable against thermal and quantum fluctuations, remains an open question even for this simple system.

\begin{figure}
	\includegraphics{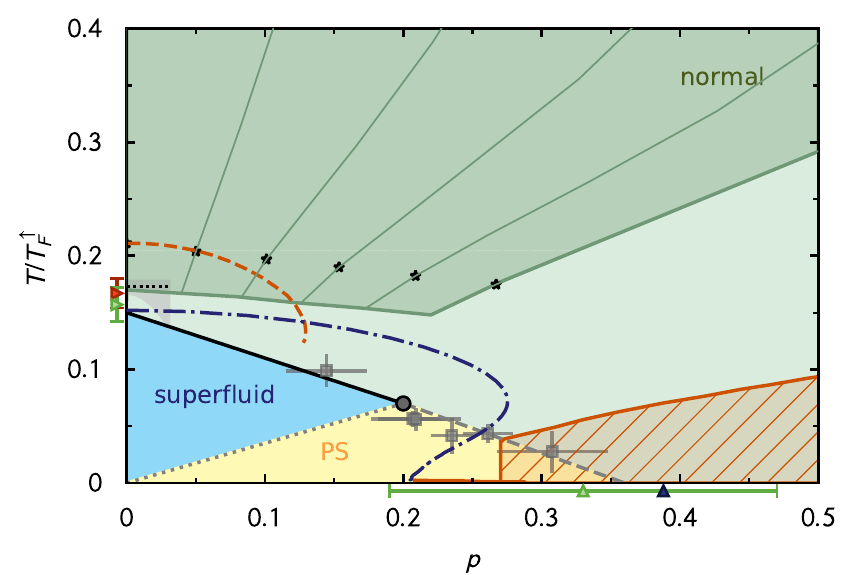}
	\caption{{\bf Phase diagram of the unitary Fermi gas} in the ($T/T_F$, $p$) plane, extended from Ref.~\cite{Rammelmuller:2021oxt}. 
	The dark green hatched region represents the domain investigated in this work. The orange hatched area marks where a recent Luttinger-Ward study 
	predicts dominant FFLO-type pairing fluctuations~\cite{pini2021strong}. Stars mark parameter values of the data sets presented below. For a detailed 
	discussion of this phase diagram, see Ref.~\cite{Rammelmuller:2021oxt}.}
	\label{fig:pd}
\end{figure}

As the formation of a superfluid condensate necessarily requires fermion pairing, the observation of a dominance in a specific channel in correlation functions can 
be viewed as a precursor for the formation of a corresponding condensate, which may even be spatially inhomogeneous.
To address the issue of pairing, suitable four-point correlation functions need to be investigated.  So far, several correlation functions have been analyzed with a variety of non-perturbative continuum methods in different approximations~\cite{Roscher:2015xha,Frank_2018,pini2021strong}, to gain a better understanding of the finite-temperature phase diagram of the spin-imbalanced unitary Fermi gas (UFG), summarized in Fig.~\ref{fig:pd}.
In these studies, the pair-correlation function -- associated with the propagator of the pairing field -- 
plays an important role as distinct maxima in the momentum-space representation  
of this correlator are expected to herald the formation of a Fulde-Ferrell-Larkin-Ovchinnikov (FFLO) ground state~\cite{pini2021strong}, i.e.,
an inhomogeneous ground state with a spatially oscillating order parameter.

In this work, we aim to shed further light on the pairing mechanisms of the polarized unitary Fermi gas by approaching the problem from the high-temperature regime. 
To this end, we track pairing through the aforementioned pair-momentum distribution as well as density-density correlations in momentum space, 
also known as the shot noise~\cite{PhysRevA.70.013603}.
Whereas the pair-momentum distribution provides us with information on the propagation of potentially condensing pairs of spin-up and spin-down fermions, 
the shot noise gives a ``tomographic'' view of the structure of the fermion pairs. In fact, 
it highlights which ``spots" in momentum-space are (anti-)correlated with each other. In this respect, the shot-noise correlator may be regarded as the covariance 
matrix of the momentum distributions of the two species.

In a similar way to earlier studies of one-dimensional setups, where the shot noise very clearly revealed dominant and even sub-dominant pairing patterns for spin-and mass-imbalanced Fermi gases~\cite{Luescher2008,rammelmueller2020}, we combine insights from both correlation functions to further map out the regions where off-center pairing 
associated with FFLO-type phases may or may not be a relevant mechanism in the three-dimensional UFG.
Strikingly, noise correlations are accessible in experiments as was recently demonstrated for two-dimensional Fermi gases, where the shot-noise correlation function was reconstructed via time-of-flight measurements of the spin-selective momentum distributions~\cite{holten2021observation}. 

%

{\it Model and Method.--}
Two-component fermions in the unitary limit are characterized by the following Hamiltonian:
\beq
 \hat{H} =\int {\rm d}^3 r\, \hat{\psi}_s^{\dagger} (\mathbf{r}) \left (- \frac{\hbar \nabla^2}{2 m } \right)  \hat{\psi}_s (\mathbf{r})\nn\\
  \!-\! g \int {\rm d}^3 r ~ \hat{n}_{\uparrow}(\mathbf{r}) \hat{n}_{\downarrow}(\mathbf{r})\,,
\eeq
where summation over~$s$ is assumed ($s=\,\uparrow,\downarrow$). Here,
$\hat \psi^{\dagger}_s({\bf r})$ and $ \hat \psi^{}_s({\bf r})$ denote
the creation and annihilation operators for fermions of spin $s$ at position $\bf r$. 
On the microscopic level, the fermion interactions assume a simple form and
can be written in terms of the density operators~$\hat n_{\uparrow,\downarrow}^{}({\bf r}) = \hat \psi^{\dagger}_{\uparrow,\downarrow}({\bf r}) \hat \psi^{}_{\uparrow,\downarrow}({\bf r})$. 
The interaction strength is controlled by the (bare) coupling parameter~$g$.
In the following, we shall set~$m=1$ for the fermion masses and also $\hbar = k_B =1$, which fixes the units in our calculations.

In this work, we shall restrict ourselves to the dilute limit, such that a zero-range interaction in
the above Hamiltonian is appropriate. Moreover, we shall only consider the so-called unitary limit which
requires to tune the system to resonance. On a finite space-time lattice, this implies a tuning of the
bare coupling, see Refs.~\cite{PhysRevC.78.024001,Lee:2008fa,PhysRevC.73.015201,Drut:2012tg} for details.

In the unpolarized many-body regime, the system described by the above Hamiltonian displays a low-temperature superfluid phase,
continuously connecting the weak-coupling (BCS) and strong-coupling (BEC) regimes, which is the well-known BCS-BEC crossover~\cite{Zwerger2012}.
As the polarization is turned on (measured by a particle number or chemical potential difference between the two spin species),
the superfluid phase is expected to shrink and eventually disappear at a critical, temperature-dependent polarization. Intriguingly, the structure
of the ground state may change from a homogeneous superfluid to a inhomogeneous superfluid state, also called supersolid,
when the polarization is increased~\cite{Chevy2010,Radzihovsky2010,Zwerger2012,GUBBELS2013255,Roscher:2015xha}.

To obtain the relevant correlation functions required to shed light on the pairing mechanism at work, we employ the complex Langevin (CL) method~\cite{Parisi:1980ys,Klauder:1983nn,Klauder:1983zm,Klauder:1983sp,Parisi:1984cs}. This non-perturbative numerical approach circumvents the sign-problem that arises in the case of spin-polarization in conventional Monte Carlo approaches. In the following, we shall only recapitulate the most important aspects of this approach; for details on the method and the sign problem we refer to Refs.~\cite{Berger:2019odf,attanasio_complex_2020}.

Our starting point is the grand-canonical partition function, given by
\beq
	\CZ = {\text{Tr}}\left[e^{-\beta (\hat{H} - \mu_\up \hat{N}_\up - \mu_\down \hat{N}_\down)}\right] \,,
\eeq
where $\beta$ is the inverse temperature, $\mu_s$ is the chemical potential for spin-$s$ particles, and $\hat N_s$ is the corresponding
particle number operator. Using a Suzuki-Trotter factorization, which defines an imaginary-time lattice
of spacing $\tau$ and extent $N_\tau$ (such that $\beta = \tau N_\tau$), followed by a density-channel Hubbard-Stratonovich (HS)
transformation, one obtains the following path integral form of the partition function:
\beq
\CZ = \int \CD \sigma \det M_\up [\sigma] \det M_\down [\sigma]\,.
\eeq
Here,~$\sigma$ is a real-valued spacetime varying HS field and $M_s$ is the Fermi matrix associated with spin-$s$ particles.
Whereas for unpolarized systems $\det M_\up \det M_\down = \det M_\up^2$ is real and non-negative since $M_s$ is real-valued (i.e., $\mu_{\uparrow}=\mu_{\downarrow}$),
there is no guarantee that the product of the determinants is non-negative at finite polarization,
which results in one of the most important open problems in quantum many-body physics across all areas: the aforementioned sign problem.
This implies that ordinary importance-sampling-based quantum Monte Carlo methods, using the determinant product as a probability measure,
only work in particular situations.

Instead of importance sampling with a repeated accept/reject step, the CL method obtains a properly distributed collection of HS fields by solving the appropriate set of two Langevin equations
\beq
\frac{d\sigma}{dt} = - \frac{\delta S[\sigma]}{\delta \sigma} + \eta\,,
\eeq
where $\sigma$ has been promoted to a complex quantity. Here, $t$ is a fictitious time parametrizing the configuration space trajectory, $S[\sigma] = -\ln (\det M_\up [\sigma] \det M_\down [\sigma])$
is the effective action, and $\eta$ represents a (real) white noise term with vanishing autocorrelation.
Although there are potential caveats to this approach~\cite{Aarts:2017vrv,nagata_argument_2016,scherzer_controlling_2020}, this strategy has been successfully applied to strongly-coupled ultracold matter~\cite{PRD96094506,2018UFGviaCL,rammelmueller2020,Attanasio2020,Rammelmuller:2021oxt,PhysRevResearch.3.033180} and thus is expected to faithfully represent the correlation functions of interest for this work. Our numerical setup and (hyper-)parameters are as in Ref.~\cite{2018UFGviaCL} with a spacetime lattice size of $V\times N_\tau=11^3 \times 160$, 
sufficient to study the UFG down to temperatures slightly below the superfluid phase transition.

{\it Results.--}
\begin{figure}[t]
	\includegraphics[width=\columnwidth]{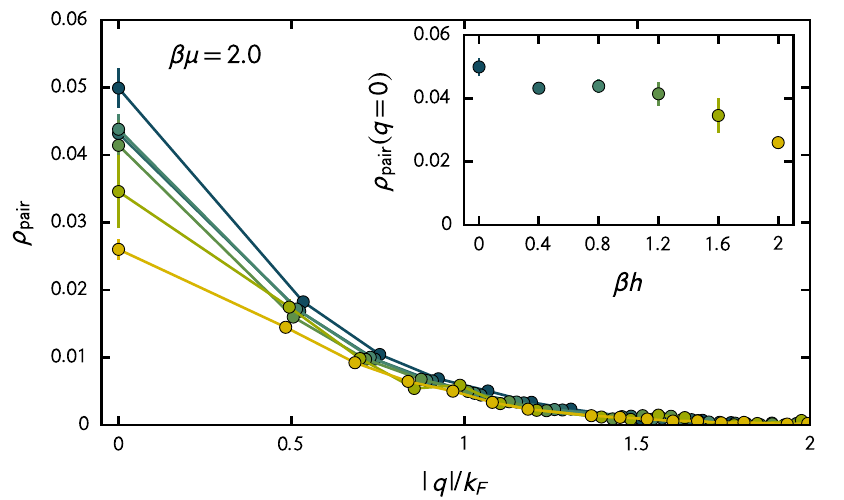}
	\caption{
		{\bf Pair-momentum distribution} $\rho_{\rm pair}$ at fixed $\beta\mu = 2.0$ for $\beta h=0 \dots 2.0$.
		(Inset) Zero-momentum component $\rho_{\rm pair}(q = 0)$ as function of the chemical potential mismatch $\beta h$.
	}
	\label{fig:pairmomdist}
\end{figure}
In this work, we aim at an investigation of the pairing mechanisms in spin-polarized unitary Fermi gases. In terms of spin polarization, we have restricted ourselves to the regime $\beta h < 2.0$, where~$h=\mu_{\uparrow}-\mu_{\downarrow}$ is the so-called Zeeman field.
This corresponds to the polarizations~$p=(n_\up - n_\down)/(n_\up + n_\down)$ marked with the dark green hatched area in Fig.~\ref{fig:pd}. With respect to the temperature, we focus on the regime above the superfluid phase transition temperature, i.e., ~$\beta \mu \leq (\beta \mu)_{\rm c}$, where~$(\beta \mu)_{\rm c} \approx 2.5$ is the phase transition temperature of the balanced system~\cite{Ku2012,Boettcher2014,Frank_2018,PhysRevB.99.094502,PhysRevLett.125.060403}. In units
of the Fermi temperature of the spin-up fermions, this corresponds to the temperature regime~$T/T^{\up}_{F} > (T_{\rm c}/T^{\up}_{F}) \approx 0.16$. Note that the phase transition temperature
is generally expected to decrease with increasing polarizations~\cite{Chevy2010,GUBBELS2013255,Boettcher2014,Frank_2018}. 
The largest polarization considered in the present work is~$p\approx 0.5$ for $T/T^{\up}_{F}\approx 0.29$.

Let us now turn to the correlation functions. The $\up$-$\down$ two-body density is defined as 
\beq
	\rho_{2,\up,\down}({\bf x}',{\bf x},{\bf y}',{\bf y}) =
	\langle
		\hat \psi^\dagger_\up({\bf x}')
		\hat \psi^\dagger_\down({\bf y}')
		\hat \psi^{}_\down({\bf y})
		\hat \psi^{}_\up({\bf x})
	\rangle.
	\label{eq:tbdm}
\eeq
To study pair correlations across a distance $\bf r$, we simply set ${\bf x}' = {\bf x} + {\bf r}$ and ${\bf y}' = {\bf y} + {\bf r}$.
The relative positions of the particles in the pair is ${\bf \tilde{r}}= {\bf y}' - {\bf x}' = {\bf y} - {\bf x}$ and the initial center-of-mass location is ${\bf R} = ({\bf x} + {\bf y})/2$. As the latter is arbitrary, one may average over it, leaving two variables ${\bf r}$ and ${\bf \tilde{r}}$ characterizing the pair correlations.
To obtain the pair-momentum distribution~$\rho_{\text{pair}}$ of tightly bound (on-site) $\up\down$-pairs, 
we set ${\bf \tilde{r}} = 0$ and compute the Fourier transform of~Eq.~(\ref{eq:tbdm}) with respect to $\bf r$. 
For the normalization of~$\rho_{\text{pair}}$, we choose
\beq
	\frac{1}{V}\int \frac{{\rm d}^3 q}{(2\pi)^3}\, \rho_{\text{pair}}({\bf q}) = n_{\uparrow}n_{\downarrow}\,,
\eeq
where~$V$ denotes the spatial volume. Thus, $\rho_{\text{pair}}$ is normalized with respect to the total number of possible combinations of $\up$- and $\down$-fermions.
\begin{figure*}[t]
	\includegraphics[width=\textwidth]{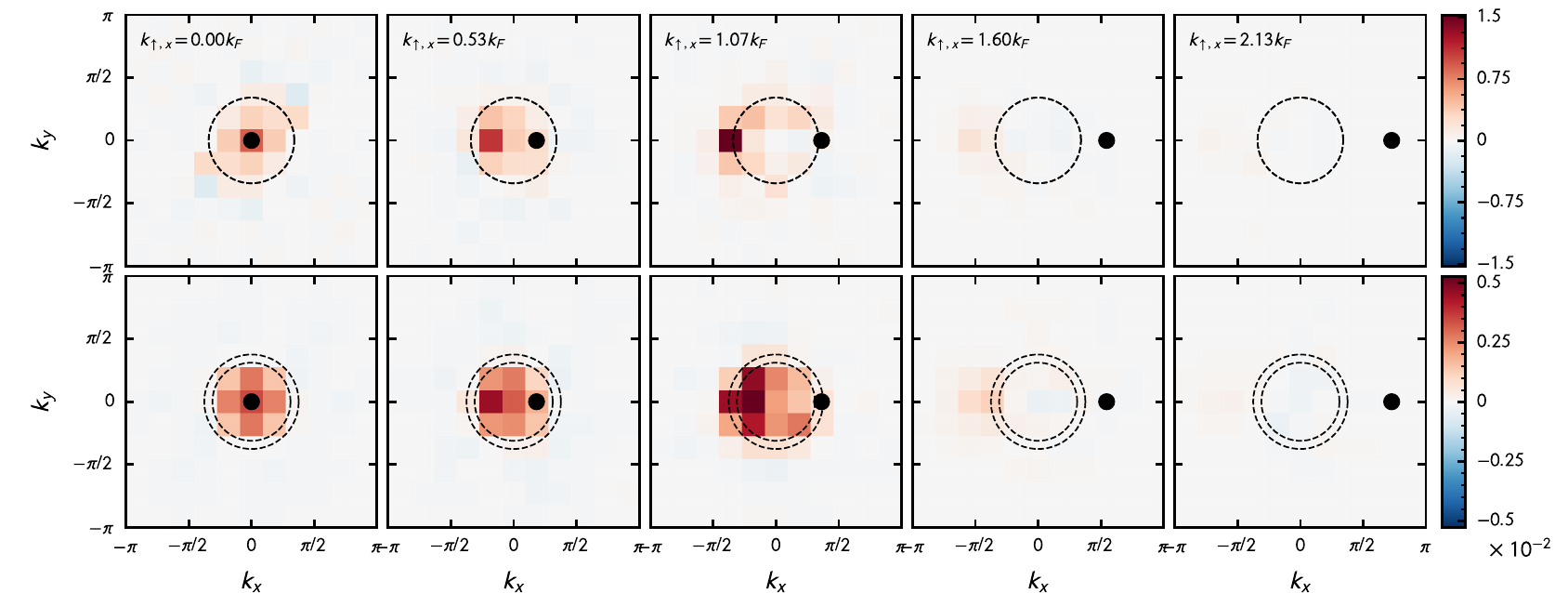}
	\caption{{\bf Density-density correlation in momentum space} at fixed $\beta\mu = 2.0$ as well as $\beta h = 0.0$ (top row) and  $\beta h = 2.0$ (bottom row) in the plane where $k_z = 0$.
	Different panels correspond to different reference positions of an $\up$-particle along the line $k = (k_x, 0, 0)$, indicated by a black dot. Dashed lines correspond to the respective Fermi surfaces.}
	\label{fig:shotnoise_2}
\end{figure*}

As mentioned above, the pair-momentum distribution has already been employed in the past to search for FFLO-type pairing, as a peak in this quantity
at ${\bf q} \neq 0$ points to unconventional pairing. On the other hand, a pair-momentum correlation peaked around ${\bf q} = 0$ is indicative of conventional BCS pairing. 
In Fig.~\ref{fig:pairmomdist}, our results for the pair-momentum distribution evaluated at~$\beta\mu =2.0$ (i.e., close to the superfluid phase transition) for various values of~$\beta h$ are shown. We observe that this distribution indeed develops a clear maximum. However, for all temperatures and polarizations considered in this work, we only find a maximum at~${\bf q}=0$ which should be viewed as a precursor for the formation of a 
conventional BCS-type superfluid ground state at low temperatures.
Moreover, we find that the height of the maximum of~$\rho_{\text{pair}}$ decreases when the polarization is increased for a fixed value of~$\beta\mu$, which is
in accordance with the fact that the phase transition line decreases for increasing polarization. Thus, for a fixed value of~$\beta\mu$, the system is farther away from the phase transition for larger polarizations and pairing correlations become weaker. In any case, there is no indication of an FFLO-type pairing signature in our results for the pair-momentum distribution, in the entire regime of temperatures and polarizations considered in this work, as illustrated in Fig.~\ref{fig:pd}.

As the pair-momentum distribution does not allow to gain an insight into the internal momentum structure of the two fermions forming a pair, we extract this important aspect from the aforementioned shot-noise correlations. In fact, even if the pair-momentum distribution indicates that the formation of pairs with a given momentum~${\bf q}$ is favored, we do not know {\it a priori} whether
the pair is formed out of, e.g., two fermions with opposite momenta of the order of the respective
Fermi momentum of the two species. Whereas this is likely to be the case for unpolarized systems,
less is known about this aspect in the presence of a finite polarization.
In fact, there is an infinite number of possible configurations for a given center-of-mass momentum of the pair. All of them may then build up a maximum in the pair-momentum distribution which may eventually be interpreted as a certain type of pairing mechanism, e.g.,
conventional FFLO-type pairing.
For one-dimensional Fermi gases with a finite spin and mass polarization, it has indeed been recently found that unconventional pairing patterns in terms of
the internal momentum structure of the pairs may be more likely than the conventional FFLO-type pattern~\cite{rammelmueller2020}.

The momentum shot-noise correlation across different spin species (not to be confused with the Fourier transform of the density-density correlation in
coordinate space, which would give the structure factor) is given by
\beq
	G_{\up,\down}({\bf k}, {\bf k}') =
	\langle \hat n_\up({\bf k}) \hat n_{\down}({\bf k}') \rangle - \langle \hat n_\up({\bf k}) \rangle \langle \hat n_{\down}({\bf k}') \rangle\,.
\eeq
Here,~${\bf k}$ and~${\bf k}^{\prime}$ refer to momenta and~$\hat n_s({\bf k})$ is the momentum-space
representation of the density operator of the fermion with spin~$s$.

\begin{figure}[t]
	\includegraphics{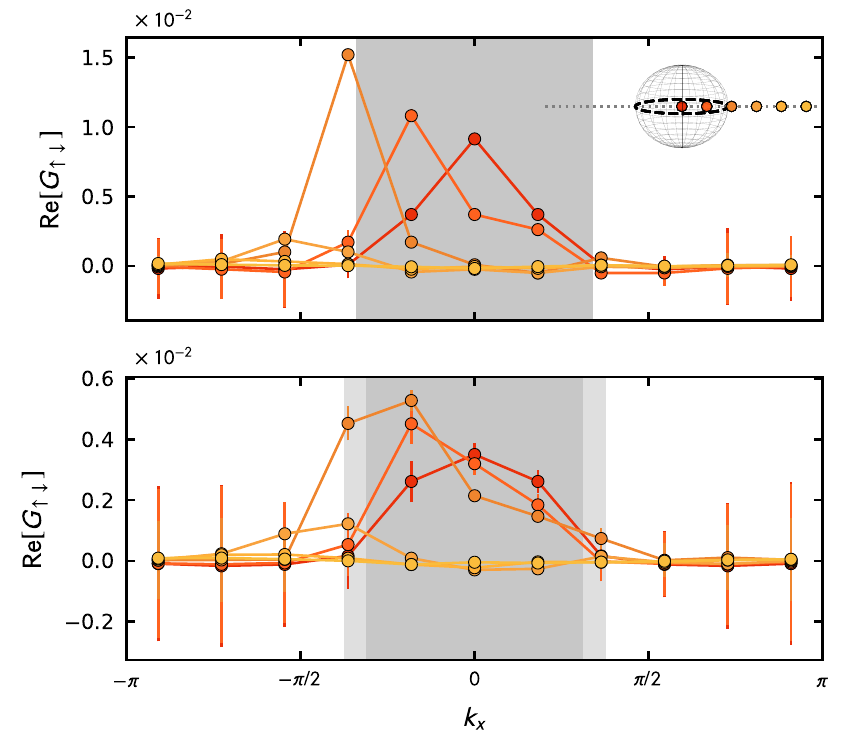}
	\caption{{\bf Cuts along $k_x$ of the shot noise in momentum space} for $\beta\mu = 2.0$ (top panel) and $\beta h = 2.0$ (bottom panel). 
	The color-coding corresponds to the reference positions indicated in the inset. The cuts correspond to the central horizontal bins in Fig.~\ref{fig:shotnoise_2}.
	The Fermi seas of the two species are indicated by gray-shaded areas. Note that, for~$\beta h >0$, the size of the two Fermi surfaces differ.}
	\label{fig:shotnoise_1}
\end{figure}

Our results for the shot-noise correlations at $\beta\mu=2.0$ are shown in Fig.~\ref{fig:shotnoise_2} for the unpolarized case ($\beta h =0$, top row) and for the polarized case ($\beta h =2.0$, bottom row). 
For better visualization, we present these correlations in the $(k_x,k_y)$-plane at $k_z=0$ for a test particle kept fixed at a given momentum in the same plane, which is 
indicated by the black dot in the various panels. For a detailed comparison, we additionally show cuts of the same quantity along the~$k_x$ axis in Fig.~\ref{fig:shotnoise_1} (corresponding to the central horizontal bins in Fig.~\ref{fig:shotnoise_2}).

In the balanced case, we observe the strongest correlation between fermions with opposite momenta ``sitting" right at the Fermi surface, as also expected for a conventional BCS superfluid. Moreover, we find that the correlation decreases when we consider such pairs with momenta smaller than the Fermi momentum.
Thus, the Fermi sea is dismantled starting from the Fermi surface down to the interior of the Fermi sea, eventually leading to the formation of a superfluid condensate of compound bosons composed of one spin-up and one spin-down fermion at sufficiently low temperatures. In particular, the shot-noise correlations suggest that the formation of pairs with fermions coming with
opposite momenta (but zero total momentum) is energetically most favorable. Intriguingly, increasing the polarization, we observe that this clear pairing pattern is washed out and other pairing channels appear to open up. In particular, we find that pairing channels with a non-vanishing center-of-mass momentum become
increasingly favorable, although such patterns (e.g., FFLO-type maximum) are not yet visible in the pair-momentum distribution. This behavior may be viewed as a precursor for the formation of unconventional condensates at even higher polarizations, as suggested in a recent $T$-matrix study~\cite{pini2021strong}. In any case, our present
{\it ab initio} study suggests that there is no indication for the formation of inhomogeneous ground states emerging from unconventional pairing patterns in the dark green hatched area in Fig.~\ref{fig:pd}.

%

{\it Summary.--}
In the present work, we have used the CL method to elucidate the onset of pairing correlations in the polarized regime of the unitary Fermi gas at high temperatures over a wide range of
polarizations. To that end, we analyzed the pair-momentum distribution and shot-noise correlations. Based on both of these, we are able to place non-perturbative,
{\it ab initio} bounds on the possible locations of exotic pairing phases such as FFLO. More precisely, for the polarizations considered in this work (cf. Fig.~\ref{fig:pd}), 
we found only indications for the formation of a conventional superfluid as indicated
by standard BCS-type pairing. However, while increasing the polarization, the shot-noise correlations suggest that the clear BCS signature is washed out and other pairing
channels become increasingly favorable, which could be viewed as the precursor for the formation of inhomogeneous condensation at even higher polarizations.
Still, the pair-momentum distribution exhibits a peak at vanishing momentum, indicating that the formation of a homogeneous superfluid ground is most favored in the range of polarizations considered here.
Of course, these findings do not exclude the existence of an inhomogeneous phase at larger polarizations and lower temperatures. However, an {\it ab-initio} analysis of this regime requires further developments of our CL framework which is deferred to future work.

Finally, we emphasize that our results for the pair-momentum distribution as well as the shot-noise correlations represent experimentally accessible predictions.
In fact, shot-noise correlations have been just recently measured in two-dimensional Fermi gases~\cite{holten2021observation}. 
We expect that an extension of such experiments to three-dimensional 
systems analyzed in the light of corresponding results from theoretical studies will further push our understanding of pair formation and 
condensation in strongly correlated systems under extreme conditions of temperature under polarization.

{\it Acknowledgments.--}
The authors thank F.~Ehmann for useful discussions. 
The work of F.A. was supported by the Deutsche Forschungsgemeinschaft (DFG, German Research Foundation) under Germany's Excellence 
Strategy EXC2181/1 - 390900948 (the Heidelberg STRUCTURES Excellence Cluster) and under the Collaborative Research Centre SFB 1225 (ISOQUANT).
L.R. is supported by FP7/ERC Consolidator Grant QSIMCORR, No.~771891, and the DFG 
under Germany's Excellence Strategy -- EXC--2111--390814868.
This material is based upon work supported by the National Science Foundation under Grants No. PHY1452635 and No. PHY2013078.
J.B. acknowledges support by the DFG under grant BR~4005/5-1 as well as under grants BR~4005/4-1 and BR~4005/6-1 (Heisenberg program).
Part of the numerical calculations have been performed on the LOEWE-CSC Frankfurt.


%
\bibliography{cold_gases}

\end{document}